\documentclass[10pt,twocolumn]{article}
\usepackage{latex8}
\usepackage{acronym}
\usepackage{psfig}
\usepackage{epsfig}
\usepackage{side}
\usepackage{html}
\usepackage{fancyhead}

\ifx \htmlstyloaded\relax  \else\let\htmlstyloaded\relax\fi

\newcommand{\htmladdnormallink}[2]{#1}

\newcommand{\htmladdimg}[1]{}

\newcommand{\externallabels}[2]{}

\newcommand{\externalref}[1]{}

\makeatletter
\def\makeinnocent#1{\catcode`#1=12 }
\def\csarg#1#2{\expandafter#1\csname#2\endcsname}

\def\ThrowAwayComment#1{\begingroup
    \def\CurrentComment{#1}%
    \let\do\makeinnocent \dospecials
    \makeinnocent\^^L
    \endlinechar`\^^M \catcode`\^^M=12 \xComment}
{\catcode`\^^M=12 \endlinechar=-1 %
 \gdef\xComment#1^^M{\def\test{#1}
      \csarg\ifx{PlainEnd\CurrentComment Test}\test
          \let\html@next\endgroup
      \else \csarg\ifx{LaLaEnd\CurrentComment Test}\test
            \edef\html@next{\endgroup\noexpand\end{\CurrentComment}}
      \else \let\html@next\xComment
      \fi \fi \html@next}
}
\makeatother

\def\includecomment
 #1{\expandafter\def\csname#1\endcsname{}%
    \expandafter\def\csname end#1\endcsname{}}
\def\excludecomment
 #1{\expandafter\def\csname#1\endcsname{\ThrowAwayComment{#1}}%
    {\escapechar=-1\relax
     \csarg\xdef{PlainEnd#1Test}{\string\\end#1}%
     \csarg\xdef{LaLaEnd#1Test}{\string\\end\string\{#1\string\}}%
    }}

\excludecomment{comment}

\excludecomment{rawhtml}

\excludecomment{htmlonly}
\newcommand{\html}[1]{}

\newcommand{\hyperref}[4]{#2\ref{#4}#3}

\newcommand{\htmlimage}[1]{}

\newcommand{\htmladdtonavigation}[1]{}

\newcommand{\aparms}{The parameters are chosen to accommodate the
mobile wireless network application described in \cite{BushThesis}.
Virtual messages are injected by the \acl{GPS} at a rate of 0.03
per millisecond ($\lambda_{vm} = 0.03$) with a lookahead of 30.0
milliseconds ($\Delta_{vm} = 30.0$). The expected time to create a
beam table is 7.0 milliseconds ($\tau_{task} = 7.0$). The expected
rollback time is 1.0 milliseconds ($\tau_{rb} = 1.0$) and the speedup in
reading from cached results over computing the beam table is 100
($C_r=100$)}

\begin{document}
%
%
\title{Active Virtual Network Management Protocol}
\author{\htmladdnormallink{Stephen F. Bush}{http://www.crd.ge.com/\~{}people/bush}\\
General Electric Corporate Research and Development\\
KWC-512, One Research Circle, Niskayuna, NY 12309\\
bushsf@crd.ge.com (http://www.crd.ge.com/people/bush)}
\maketitle
\thispagestyle{empty}
\begin{abstract}

This paper introduces a novel algorithm, the \ac{AVNMP},
for predictive network management. It explains how 
the \acl{AVNMP} facilitates the management of an active network by 
allowing future predicted state information within an active network 
to be available to network management algorithms. This is
accomplished by coupling ideas from optimistic discrete event 
simulation with active networking. The optimistic discrete event 
simulation method used is a form of self-adjusting Time Warp. It
is self-adjusting because the system adjusts for predictions which are
inaccurate beyond a given tolerance. The concept of a streptichron and 
autoanaplasis are introduced as mechanisms which take advantage of the 
enhanced flexibility and intelligence of active packets. Finally, it is 
demonstrated that the \acl{AVNMP} is a feasible concept.

\end{abstract}


%
%

\newif\ifisdraft
\isdraftfalse

\Section{Network Management and Active Networks}

The problem this paper addresses is the complexity of managing
large and rapidly growing communication networks. Network management 
consists of a wide variety of responsibilities including configuration 
management, performance management, fault management, accounting management, 
and security management. A network management system must be able to monitor,
control, and report upon the status of all of these areas. This is
usually performed using a standards based management protocol such as
the \acf{CMIP} \cite{CMIP} or the \acf{SNMP} \cite{SNMP}. A goal of network
management is to pro-actively detect problems in each of these areas. This
means detecting such events as performance problems and faults before
they occur. This is accomplished by the \acl{AVNMP}.

Active networks \cite{TSSWM97} are a relatively recent concept in 
communication networks. Active networks are capable of executing 
general purpose code within packets as the packets are transmitted 
through intermediate network nodes. A framework for supporting the
execution of general purpose code within packets as they travel through
a network is an on-going research effort. Thus active networks differ from 
today's communications networks because active networks offer a computational 
service in addition to a data transport service. 
In current communication networks non-executable data is passively
forwarded through the traditional communication layers; intermediate
devices such as bridges and routers only access the data link or network
headers of packets. In active networks, intermediate devices can execute
generic code within active packets as they travel through the network.
The ability for communication networks to perform such computation offers 
opportunities for great advantages in such areas as efficiency, rapid 
protocol development and deployment, and network flexibility. 
However, active networks also add additional complexity, particularly in 
network management and security. The goal of the \acl{AVNMP} is to use the 
advantages active networks provide in order to handle the additional 
complexity in network management.

\begin{figure}[htbp]
	\centerline{\psfig{file=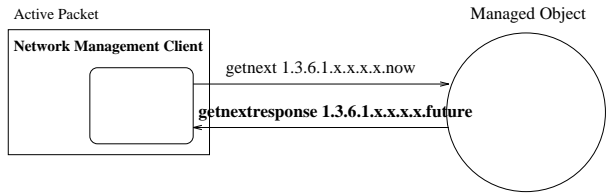,width=3.125in}}
        \caption{Obtaining a Future MIB Object Value.}
        \label{ft}
\end{figure}

The \acl{AVNMP} caches predicted values within a State Queue and makes 
them available to a standard network management interface such as 
the \ac{SNMP} \cite{SNMP} as shown in Figure \ref{ft}. Because time is 
appended to the Object Identifier, 
a series of \textbf{Get-Next} requests will return all the predicted 
cached values of a \ac{MIB} object. Also note in Figure \ref{ft} that 
the \ac{SNMP} agent has the capability to reside within a packet. Because 
it is an active network, packets are capable of issuing management requests 
and responding to such requests.

The predictive capability provided by the \acl{AVNMP} facilitates the 
development of a variety of predictive applications from mobile wireless
location management and network security to improved \ac{QoS}.
Mobile systems, especially those using the \acl{GPS} can predict their
location. This information can be propagated through the system
via the \acl{AVNMP}. An example of predictive mobile wireless location 
management 
is described in \cite{BushWinet97}. Because the mobile host's location
can be predicted, the setup information required for hand-off can be cached 
ahead of time providing dramatic increases in the speed of hand-off and
thus improved \ac{QoS}. In the area of network security,
given a set of vulnerabilities within a communications network, the most
probable path an attacker will follow can be determined. The \acl{AVNMP} can
thus incorporate the effect of an attack in its prediction and
the consequences of a real or anticipated attack can be propagated through 
the system before it occurs. In regards to \acl{QoS},
time sensitive applications require a predictable \acl{QoS}
\cite{BrClSh94}. The offered load at the input to the network can be 
predicted and the \acl{AVNMP} will transparently propagate the predicted 
load through intermediate network devices. An example of using
predictive load management is described in \cite{BushIEEEBB96}. An example 
of a method for predicting network traffic based on Wavelets 
\cite{nips-10:Ma+Ji:1998} shows promise and can be used to implement the 
\ac{DP} within the \acl{AVNMP}. The \acl{AVNMP} driving process is
described in Section \ref{dp}.
\Section{Active Virtual Network Management Protocol Description}
\label{vncorg}

\begin{figure*}[htbp]
        \centerline{\epsfig{file=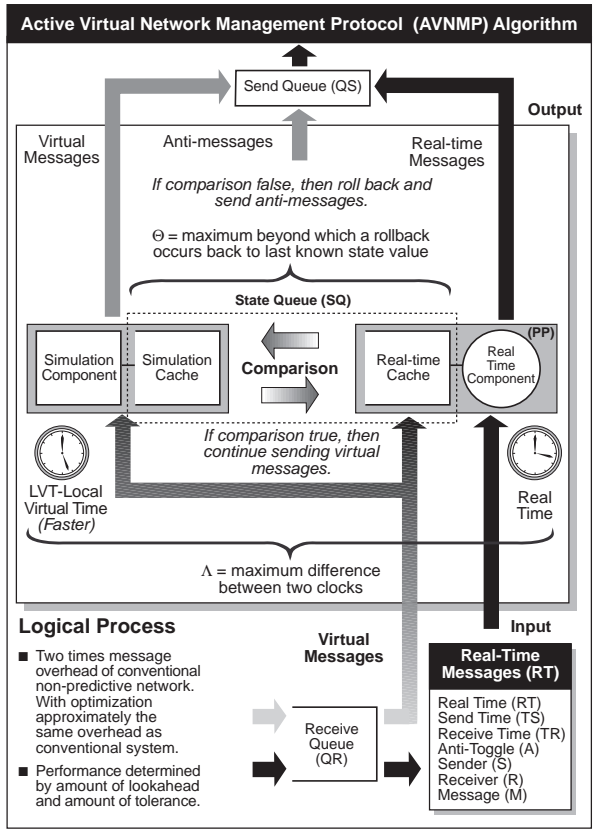,width=3.125in}}
        \caption{The AVNMP Logical Process.}
        \label{lpdet}
\end{figure*}

The \acl{AVNMP}\index{AVNMP} algorithm encapsulates each \acl{PP}\index{PP} 
within a \acl{LP}\index{LP} as illustrated in Figure \ref{lpdet}. 
A \acl{PP}\index{PP} is nothing more than an executing task implemented
by program code. An example of a \acl{PP} in a mobile wireless environment
is the \acl{RDRN} \cite{RDRN} beam table computation task. The beam table 
computation task generates a table of complex weights which controls the 
angle of radio beams based on position input.
A \acl{LP}\index{LP} consists of the \acl{PP}\index{PP} and additional 
data structures and instructions which maintain message 
order and correct operation as the system executes ahead of real 
time\index{AVNMP!real time}. These structures are illustrated in detail
in Figure \ref{lpdet}. As an example, the beam table computation
\acl{PP} is encapsulated in a \acl{LP} which maintains generated beam
tables in its State Queue and handles rollback due to 
out-of-order input messages or out-of-tolerance real messages as
explained in Section \ref{rb}. A \acl{LP}\index{LP}
contains a Receive Queue (QR\index{QR|see{Receive Queue}}),
Send Queue\index{Send Queue} (QS\index{QS|see{Send Queue}}), and 
State Queue\index{State Queue} (SQ\index{SQ|see{State Queue}}) as shown
in Figure \ref{lpdet}. The simulation component and simulation cache on
the left side of Figure \ref{lpdet} represent the execution and state
based upon virtual messages. The real time component and cache represent 
the execution and state based upon real messages.
The \acl{QR} maintains newly arriving messages in order by their Receive 
Time\index{Receive Time} (TR\index{TR|see{Receive Time}}). 
The \acl{QS} maintains copies of previously sent messages in order of 
their send times. The state of a \acl{LP}\index{LP} is
periodically saved in the \acl{SQ}. The \acl{LP}\index{LP} also contains 
its notion of time known as Local Virtual Time\index{Local Virtual Time}
(LVT\index{LVT|see{Local Virtual Time}}) and a 
Tolerance\index{Tolerance} ($\Theta$\index{$\Theta$|see{Tolerance}}).
The tolerance is the allowable deviation between actual and predicted
values of incoming messages. For example, when a real message enters the 
beam table computation \acl{LP} the position in the message value is compared
with the position which had been cached in the State Queue of the
\acl{LP}. If these values are out of tolerance, then corrective action
is taken in the form of a rollback as explained in Section \ref{rb}.

The \acl{AVNMP} \acl{LP} has the contents
shown in Table \ref{lpspectab}, the message fields are shown in
Table \ref{messpectab}, and the message types are listed in
Table \ref{mestypes} where $t$ is the real time at the receiving
\acl{LP}. \acl{AVNMP} messages contain the Send Time\index{Send Time} 
(TS\index{TS|see{Send Time}}), 
Receive Time\index{Receive Time} (TR\index{TR|see{Receive Time}}), 
Anti-toggle\index{Anti-toggle} 
(A\index{A|see{Anti-toggle}}) and the actual message value itself (M).
The \ac{TR} is the time this message is predicted to be valid at
the destination \acl{LP}\index{LP}. It is not the link transfer 
time from source to destination \acl{LP}.
The \ac{TS} is the time this message was sent by the
originating \acl{LP}\index{LP}. The ``A'' field is the anti-toggle and is 
used for creating an anti-message\index{Anti-Message} to remove the effects
of false messages\index{False Message} as described
later. A message also contains a field for current 
Real Time\index{Real Time}
(RT\index{RT|see{Real Time}}). This is used to differentiate a real 
message\index{Real Message} from a virtual message\index{Virtual Message}.
A message which is generated and time-stamped with the current time is
called a real message\index{Real Message}. Messages which 
contain future event information and are time-stamped with a time greater 
than current time are called virtual messages\index{Virtual Message}. 
If a message arrives at a \acl{LP}\index{LP} out of order or with invalid 
information, it is called a false message\index{False Message}. 
A false message\index{False Message} will cause a
\acl{LP}\index{LP} to rollback\index{Rollback}.

\begin{table}[htbp]
\centering
\begin{tabular}{||l|l||} \hline
\textbf{Structure} & \textbf{Description} \\ \hline \hline
Receive Queue (QR) & Messages ordered \\
                   & by receive time (TR) \\ \hline
Send Queue (QS)    & Messages ordered \\
                   & by send time (TS) \\ \hline
Local Virtual  & \\
Time (LVT) & $LVT = \inf RQ$ \\ \hline
State Queue (SQ) & States are periodically saved \\ \hline
{\em Sliding Lookahead}  & \\
{\em Window} & $SLW = (t, t + \Lambda$] \\ \hline
{\em Tolerance ($\Theta$)} & Allowable deviation \\ \hline
\end{tabular}
\caption{\label{lpspectab}\acl{AVNMP} \acl{LP} Structures.}
\end{table}

\begin{table}[htbp]
\centering
\begin{tabular}{||l|l||} \hline
\textbf{Field} & \textbf{Description} \\ \hline \hline
Send Time (TS) & LVT of sending process  \\
               & when message is sent \\ \hline
Receive Time (TR) & Scheduled time message \\
                  & is to be received \\ \hline
Anti-toggle (A) & Identifies message as \\
                & normal or anti-message \\ \hline
Message (M) & Contents of the message  \\ \hline
{\em Real Time (RT)} & GPS time message originated \\ \hline
\end{tabular}
\caption{\label{messpectab}\acl{AVNMP} Message Fields.}
\end{table}

\begin{table}[htbp]
\centering
\begin{tabular}{||l|l||} \hline
Virtual Message & $RT > t$ \\ \hline
Real Message    & $RT \leq t$ \\ \hline
\end{tabular}
\caption{\label{mestypes}\acl{AVNMP} Message Types.}
\end{table}

\SubSection{Driving Process}
\label{dp}

\begin{figure}[htbp]
	\centerline{\psfig{file=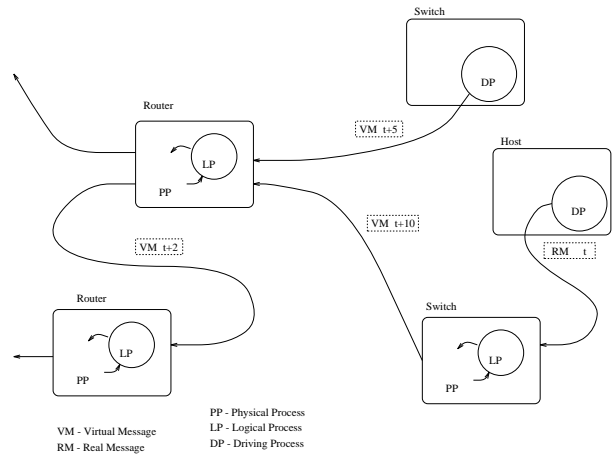,width=3.125in}}
        \caption{Driving and Logical Processes within a Communications Network.}
        \label{scope}
\end{figure}

The \acl{AVNMP} algorithm requires a \acf{DP}\index{AVNMP!driving process}
to predict future events and inject them into the system. The driving 
process acts as a source of virtual messages\index{AVNMP!virtual messages} 
for the \acl{AVNMP} system. Virtual messages are injected at a rate of
$\lambda_{vm}$ messages per unit time and each virtual message has a lookahead
of $\Delta_{vm}$ time units. \acl{LP}es react to virtual messages. For 
example, in the case of mobile wireless networking, the 
\acl{GPS}\index{Global Positioning System}
receiver process runs in real-time\index{Real Time} providing
current time and location information
and has been modified to inject future predicted time and location 
messages as well. 
Figure \ref{scope} shows an example how an active network would appear
with \acl{LP}es and \acl{DP}es deployed. Notice that the driving 
processes define the scope of the system, that is, the degree to which 
the system is predictable. For example, if only a particular route is of 
interest for load prediction purposes, then driving processes may be sent only 
to adjacent tributary nodes of the path. One of the objectives of this 
research is to enable the logical and driving processes to automatically 
and dynamically locate themselves in optimal positions within the network.

\SubSection{Rollback}
\label{rb}

A rollback is triggered either by messages arriving out of order at
the \acl{QR} of a \acl{LP} or by a predicted value previously
computed by this \acl{LP} which is beyond the allowable tolerance.
These are known as false messages because both types of messages are
sources of error.
In either case, rollback\index{AVNMP!rollback} is a mechanism by 
which a \acl{LP}\index{LP} returns to a known correct state.
The rollback\index{Rollback} occurs just as in the original Time Warp
algorithm \cite{Jefferson82}. There are three phases. In the first phase, the 
\acl{LP}\index{LP} state is restored to a virtual time strictly earlier than the 
\acl{TR} of the false message\index{False Message}. In the second phase, 
anti-messages are sent to cancel the effects of any invalid messages which 
had been generated before the arrival of the false 
message\index{False Message}. An 
anti-message\index{Anti-Message} contains exactly the
same contents as the original message with the exception of an anti-toggle
bit\index{Anti-toggle!anti-toggle bit} which is set. When the 
anti-message\index{Anti-Message} and original message meet,
they are both annihilated\index{Annihilation}. The final phase consists 
of executing the \acl{LP}\index{LP} forward in virtual time from its 
rollback\index{Rollback} state to the virtual time the 
false message\index{False Message} arrived. No messages are canceled or sent 
between the virtual time to which the \acl{LP}\index{LP} rolled back and the 
virtual time of the false message\index{False Message}. 
Because these messages are correct, there is no need to cancel or re-send 
them. This increases performance and it prevents
additional rollbacks. Note that another 
false message\index{False Message} or anti-message\index{Anti-Message} 
may arrive before this final phase has completed without causing problems.

\Section{Streptichrons}

In the \acl{AVNMP} architecture described thus far, 
there is a one-to-one correspondence between virtual messages and
real messages. While this correspondence works well for adding
prediction to protocols using a relatively small portion of the total
bandwidth, it would clearly be beneficial to reduce the message load,
especially when attempting to add prediction of the bandwidth itself. There 
are more compact forms of representing
future behavior within an active packet besides a virtual message.
For relatively simple and easily modeled systems, only the model 
parameters need be sent and used as input to the logical process on 
the appropriate intermediate device. Note that this assumes that the 
intermediate network device's \acl{LP} is simulating the device 
operation and contains the appropriate model. 
However, because the payload of a virtual message is exactly the same 
as a real message, the payload of the virtual
message can be passed to the actual device and the result from the
actual device is intercepted and cached. In this case,
the \acl{LP} is a thin layer of code between the actual device and
virtual messages primarily handling rollback.
An entire executable load model can be included within an active 
packet generated by the \acl{DP} and executed by the \acl{LP}.
When the active packet reaches the target intermediate device, the load
model provides virtual input messages to the \acl{LP} and the payload
of the virtual message passed to the actual device as previously
described.
A $\overbrace{\mbox{Strepti}}^{bend}\overbrace{\mbox{chron}}^{time}$
is an active packet facilitating prediction as shown in Definition
\ref{streptdef} which implements any of the above mechanisms.

\begin{equation}
\label{streptdef}
\mbox{Streptichron} \stackrel{\rm \Delta}{=}
  \left \{
        \begin{tabular}[l]{l}
                Input Model (Monte-Carlo) Model \\
                Model Parameters (Self Adjusting) \\
                Virtual Message (Self Adjusting) 
        \end{tabular}
 \right.
\end{equation}
\Section{Autoanaplasis}
\label{strept}

\begin{figure}[htbp]
        \centerline{\psfig{file=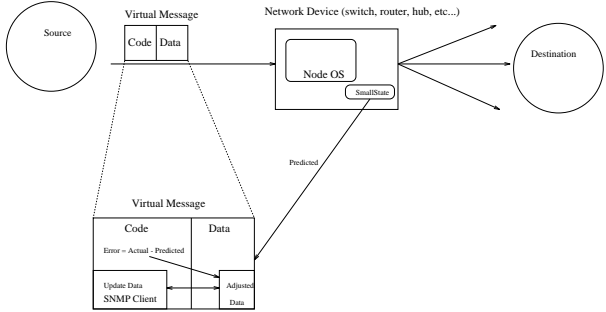,width=3.125in}}
        \caption{A Streptichron Refines its Prediction as it Travels Through the Network.}
        \label{sa}
\end{figure}

$\overbrace{\mbox{Auto}}^{self}\overbrace{\mbox{anaplasis}}^{adjusting}$
is the self-adjusting characteristic of streptichrons.
One of the virtues of the \acl{AVNMP} is the ability for the
predictive system to adjust itself as it operates. This is accomplished
in two ways. When real time reaches the time at which a predicted value
had been cached, a comparison is made between the real value and the
predicted value. If the values differ beyond a given tolerance, then the
\acl{LP} rolls backward in time. Also, active packets which implement 
virtual messages 
adjust, or refine, their predicted values as they travel through the 
network. As a specific example consider a streptichron in the form of a 
simple virtual message which anticipates load as illustrated in Figure
\ref{sa}. Although the virtual message represents a message expected to
exist in the future, the virtual message implementation exists in real
time providing feedback to
the predictive system. For example, as the  virtual message travels from one 
intermediate node to another, the packet computes its transfer time and 
compares it with the predicted state at that time of the intermediate
logical process causing a rollback if it is out of tolerance.
Also, as the streptichron travels through the network and as real
time approaches the streptichron's \acl{TR}, the streptichron
refines its predicted value. 

\Section{Class Hierarchy}

\begin{figure}[htbp]
	\centerline{\psfig{file=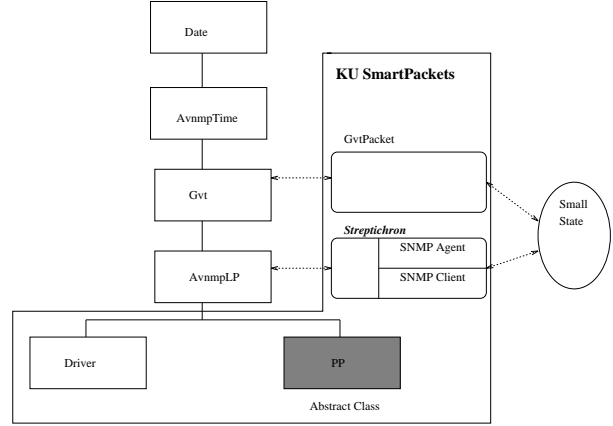,width=3.125in}}
        \caption{Active Virtual Network Management Protocol Class Hierarchy.}
        \label{code}
\end{figure}

The software architecture and state of the code under
development are shown in Figure \ref{code}. Time is critical in 
the architecture of the \acl{AVNMP} system; thus, most classes are 
derived from class \emph{Date}. Class \emph{AvnmpTime} handles relative 
time operations. Class \emph{Gvt} uses active the \emph{GvtPackets} 
class to calculate global virtual time. Class \emph{AvnmpLP} handles 
the bulk of the processing including rollback. Class \emph{Driver} 
generates and injects real and virtual messages into the system. 
The \emph{PP} class either simulates, or accesses, an actual 
device on behalf of the \acl{LP}. The \emph{PP} class may not need to 
simulate the device because the payload of a virtual message is exactly 
the same as a real message; thus, the payload of the virtual
message can be passed to the actual device and the result from the
actual device is intercepted and cached.
In this case, the \acl{LP} is a thin layer of code between the actual 
device accessed by the \emph{PP} class. The \emph{GvtPacket} class implements
the \acl{GVT} packet which is exchanged by all logical 
and driving processes to determine global virtual time. The 
\emph{Streptichron} class has been discussed in Section \ref{strept}.
Currently only the virtual message form of a streptichron has been
implemented. The active packets have been implemented in both ANTS
\cite{TSSWM97} and KU SmartPackets \cite{Kulk}.

\Section{Performance Analysis}
\label{analysis}

This section analyzes the benefit of \acl{AVNMP} in terms of
speedup based upon accurately predicting system behavior. There are 
many factors which influence speedup\index{Speedup} including
out-of-order message probability, out-of-tolerance state value
probability, rate of virtual\index{Active Virtual Network Management Protocol!verification} messages entering the system, 
task execution time, task partitioning into \acl{LP}es, rollback\index{Rollback}
overhead, prediction accuracy as a function\index{Function} of distance into the
future which predictions are attempted, and the effect of 
parallelism and optimistic synchronization. All of these factors
are considered in this section beginning with a direct analysis using the 
definitions from optimistic simulation.

\begin{figure}[htbp]
        \centerline{\psfig{file=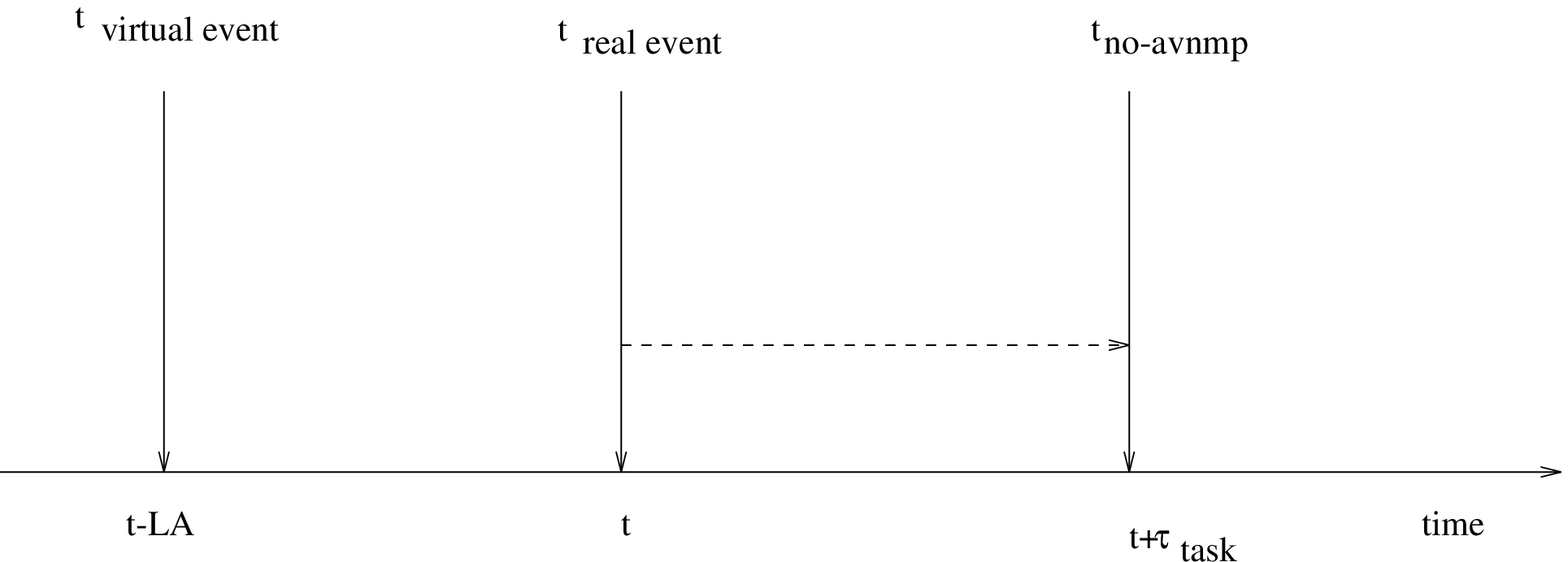,width=3.125in}}
        \caption{\acl{AVNMP} Cached before Real\index{Real Message} Event.}
        \label{cachesu1}
\end{figure}

\begin{figure}[htbp]                                                        
        \centerline{\psfig{file=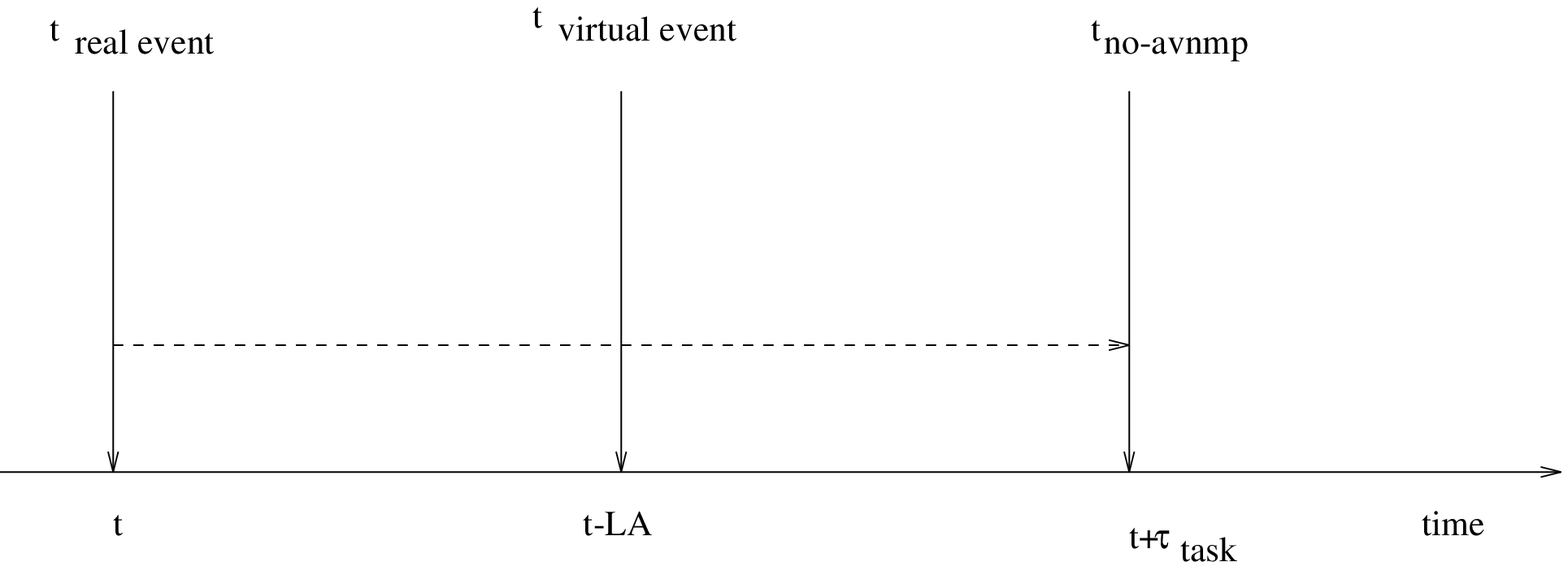,width=3.125in}}
        \caption{\acl{AVNMP} Cached later than Real\index{Real Message} Event.}
        \label{cachesu2}
\end{figure}

\begin{figure}[htbp]                                                        
        \centerline{\psfig{file=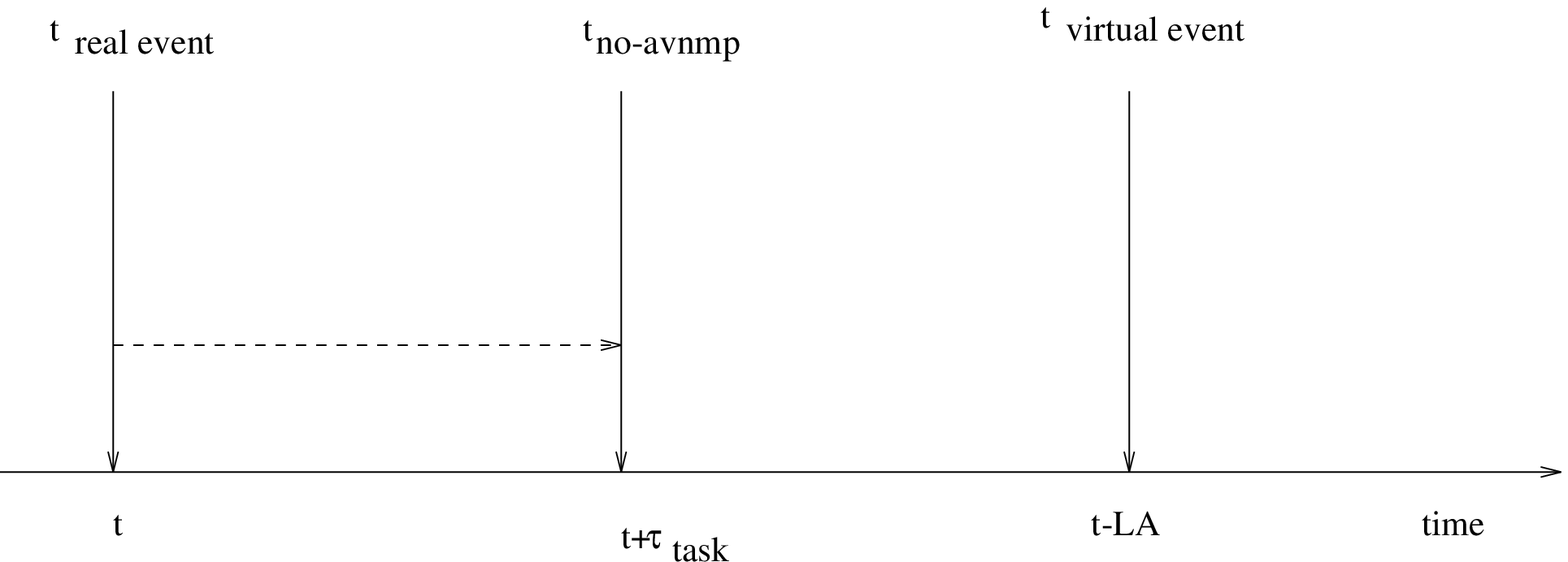,width=3.125in}}
        \caption{{AVNMP} Cached slower than Real\index{Real Message} Time\index{Time}.}
        \label{cachesu3}
\end{figure}

The definition of \acf{GVT} can be applied to determine the relationship
among expected task execution time ($\tau_{task}$), the real time at 
which the state was cached ($t_{SQ}$), and real\index{Real} time ($t$). 
Consider the value ($V_v$) which is cached at real time $t_{SQ}$ in the \acf{SQ}
resulting from a particular predicted event. 
The state queue values may be repeatedly added and discarded
as \acl{AVNMP} operation proceeds in the presence of rollback\index{Rollback}.
As rollback\index{Rollback}s occur, values for a particular predicted event may 
change, converging to the real\index{Real} value ($V_r$).
For correct operation of \acl{AVNMP}, $V_v$ should approach $V_r$ as
$t$ approaches $GVT(t_1)$ where $GVT(t_1)$ is the $GVT$ of the \acl{AVNMP} 
system at time $t_1$. Explicitly, this is for all $\epsilon > 0$ there exists 
$\delta > 0$ such that $|GVT(t_1) - t| < \delta$ implies $|f(t) - f(GVT(t_1))| 
< \epsilon$ where $f(t) = V_r$ and $f(GVT(t_1)) = V_v$. $f(t)$ is the 
prediction function\index{Function} of a driving\index{Driving Process} 
process. The purpose and function\index{Function} of the 
driving\index{Driving Process} 
process has been explained in Section \ref{vncorg}. Because \acl{AVNMP} will
always use the correct value when the predicted time ($\tau$) equals the
current real time ($t$) and it is assumed that the predictions will become
more accurate as the predicted  time of the event approaches the current time, 
the reasonable 
assumption is made that $\lim_{\tau \rightarrow t} f(\tau) = V_v$.
In order for the \acl{AVNMP} system to always look ahead, 
for all $t$ $GVT(t) \ge t$. This means that for all $n \in \{LPs\}$ and
for all $t$ $LVT_{lp_n}(t) \ge t$ and $\min_{m \in \{M\}} \{ m \} \ge t$ where
$m$ is the receive time of a message, $M$ is the set of messages in
the entire system and $LVT_{lp_n}$ is the \acl{LVT} of the $n^{th}$ \acl{LP}. 
In other words, the \acf{LVT} of each \acf{LP} must be greater than
or equal to real\index{Real} time and the smallest message \acl{TR} not yet
processed must also be greater than or equal to real\index{Real} time. The
smallest message \acl{TR} could cause a rollback to that time.
This implies that for all $n,t$ $LVT_{dp_n}(t) \ge t$. In other words,
this implies that the \acf{LVT} of each driving\index{Driving Process} 
process must be greater than or equal to real\index{Real} time. An out-of-order
rollback\index{Rollback} occurs when $m < LVT\index{LVT}(t)$. The largest saved
state time such that $t_{SQ} < m$
is used to restore the state of the \acl{LP}, where $t_{SQ}$ is the real
time the state was saved. Then the expected task execution
time ($\tau_{task}$) can take no longer than $t_{SQ} - t$ to complete
in order for $GVT$ to remain ahead of real\index{Real} time. Thus, a constraint
between expected task execution time ($\tau_{task}$), state save time
($t_{SQ}$), and real\index{Real} time ($t$) has been defined. 
There are three possible cases to consider when determining the
speedup of the \acl{AVNMP} over non-lookahead sequential
execution. In this section we will determine the speedup\index{Speedup}
given each of these cases and their respective probabilities.
These cases are illustrated in Figures \ref{cachesu1}
through \ref{cachesu3}. The time that an event is predicted to
occur and the result cached is labeled $t_{virtual\ event}$, the 
time a real\index{Real} event occurs is labeled $t_{real\ event}$, and the 
time a result for the real\index{Real} event is calculated is labeled 
$t_{no-avnmp}$. In the \acl{AVNMP}, the 
virtual\index{Active Virtual Network Management Protocol!virtual event} event and its 
result can be cached before the 
real\index{Active Virtual Network Management Protocol!real event} event as shown in 
Figure \ref{cachesu1}, between the real\index{Real} event but before the 
real event result is calculated
shown in Figure \ref{cachesu2}, or after the real event result is 
calculated as shown in Figure \ref{cachesu3}. In each case, all events
are considered relative to the occurrence of the real\index{Real} event. It
is assumed that the real\index{Real} event occurs at time $t$. A random
variable called the lookahead ($LA$) is defined as $LVT - t$. The 
virtual event occurs at time $t - LA$. Assume that the task which must
be executed once the real\index{Real} event occurs takes $\tau_{task}$
time. Then without the \acl{AVNMP} the task is completed at time
$t + \tau_{task}$.

The prediction rate of a \acl{LP} is the rate of change of the
processes' \acl{LVT} with respect to real time,
thus $PR$ is the \acl{AVNMP} speedup\index{Speedup}.
The prediction rate has been derived in \cite{BushThesis} and is
defined in Equation \ref{l+p}. 
$S_{parallel}$ is the speedup due to parallelism, $\Delta_{vm}$ is the
lookahead per virtual message, $\lambda_{vm}$ is the rate at which
virtual messages are injected by a driving process, $X$ and $Y$ are
random variables representing the proportion of out-of-order and
out-of-tolerance messages respectively, $\tau_{task}$ is the expected
amount of time taken by the physical process to process an input 
message, and $\tau_{rb}$ is the expected time taken by a \acl{LP} to
rollback. Equation \ref{l+p} includes the time to predict an event and 
cache the result in the \acf{SQ}. 

In \cite{BushThesis} the expected value of $X$ 
has been determined based on the inherent synchronization of the \acl{LP}
topology. It is shown in \cite{BushThesis} that $X$ has an expected value 
which varies with the rate of hand-offs in the mobile wireless location
management application. It is clear that the
proportion of out-of-order messages is dependent on the \acf{LP}
architecture and the partitioning of tasks into \acl{LP}es. Thus, it is
difficult in an experimental implementation\index{Active Virtual Network 
Management Protocol!implementation} to vary $X$. It is easier to change 
the tolerance rather than change the \acl{LP} architecture to evaluate
the performance of the \acl{AVNMP}. For these reasons, the analysis proceeds 
with $PR_{X,Y|X=E[X]}$.
Since the prediction rate is the rate of change of \acl{LVT} with
respect to time, the value of the \acl{LVT} is shown in Equation 
\ref{intlvt} where $C$ is an initial offset. This offset may occur 
because the \acl{AVNMP} may begin running $C$ time units before or after 
the real\index{Real} system. Replacing $LVT$ in the definition of $LA$
with the right side of Equation \ref{intlvt} yields the Equation
for lookahead shown in Equation \ref{la}.

\begin{figure*}
\begin{eqnarray}
\label{l+p}
\lefteqn{PR_{X,Y} = } & & \\
& & \lambda_{vm} \left(\Delta_{vm} S_{parallel} - \tau_{task} -
   (\tau_{task} + \tau_{rb}) X - (\Delta_{vm}S_{parallel} - {1 \over
   \lambda_{vm}}) Y\right) \nonumber
\end{eqnarray}
\end{figure*}
 
\begin{eqnarray}
\label{intlvt} \lefteqn{LVT_{X,Y|X=E[X]} = } & & \\
& \lambda_{vm} (\Delta_{vm} S_{parallel} - \tau_{task} - (\tau_{task} + \tau_{rb}) E[X] - & \nonumber \\
& (\Delta_{vm} S_{parallel} - {1 \over \lambda_{vm}}) Y) t + C & \nonumber \\
& \label{la} LA_{X,Y|X=E[X]} = (LVT_{X,Y|X=E[X]} - 1) t + C &
\end{eqnarray}

The probability of the event in which the \acl{AVNMP} result is cached
before the real\index{Real} event is defined in Equation \ref{pcache}. The probability 
of the event for which the \acl{AVNMP} result is cached after the real\index{Real} event 
but before the result would have been calculated in the non-\acl{AVNMP} system
is defined in Equation \ref{plate}. Finally, the probability of the event
for which the \acl{AVNMP} result is cached after the result would have been 
calculated in a non-\acl{AVNMP} system is defined in Equation \ref{pslow}.

\begin{eqnarray}
\label{pcache} P_{cache} = P[LA_{X,Y|X=E[X]} > \tau_{task}] \\
\label{plate} P_{late} = P[0 \le LA_{X,Y|X=E[X]} \le \tau_{task}] \\
\label{pslow} P_{slow} = P[LA_{X,Y|X=E[X]} < 0]
\end{eqnarray}

The goal of this analysis is to determine the effect of the proportion
of out-of-tolerance messages ($Y$) on the speedup\index{Speedup} of a \acl{AVNMP} system.
Hence we assume that the proportion $Y$ is a binomially distributed\index{Distributed} random 
variable with parameters $n$ and $p$ where $n$ is the total number 
of messages and $p$ is the probability of any single message being out of 
tolerance. It is helpful to simplify Equation \ref{la} by using
$\gamma_1$ and $\gamma_2$ as defined in Equations 
\ref{g1} and \ref{g2} in Equation \ref{sla}.

\begin{eqnarray}
\label{sla} LA_{X,Y|X=E[X]} = \gamma_1 - \gamma_2 Y & & \\
\label{g1} \gamma_1 = (\lambda_{vm} \Delta_{vm} S_{parallel} - \lambda_{vm} \tau_{task} - & & \\
\lambda_{vm} (\tau_{rb} + \tau_{task}) E[X]) - 1) t + C & & \nonumber \\
\label{g2} \gamma_2 = \lambda_{vm} (\Delta_{vm} S_{parallel} - {1 \over \lambda_{vm}} + \tau_{rb}) t & &
\end{eqnarray}

The early prediction probability as illustrated in Figure \ref{cachesu1} 
is shown in Equation \ref{fpcache}. The late prediction probability as
illustrated in Figure \ref{cachesu2} is shown in Equation \ref{fplate}. 
The probability for which \acl{AVNMP} falls behind real\index{Real} time as illustrated 
in Figure \ref{cachesu3} is shown in Equation \ref{fpslow}.
The three cases for determining \acl{AVNMP} speedup\index{Speedup} are thus
determined by the probability that $Y$ is greater or less than two
thresholds.

\begin{eqnarray}
\label{fpcache}P_1(t) = P_{{cache}\ {X,Y|X=E[X]}} = P\left[Y < {{\gamma_1 - \tau_{task}} \over \gamma_2}\right] & & \\
\label{fplate}P_2(t) = P_{{late}\ {X,Y|X=E[X]}} = P\left[{{\gamma_1 - \tau_{task}} \over \gamma_2} \le Y \le {\gamma_1 \over \gamma_2}\right] & & \\
\label{fpslow} P_3(t) = P_{{slow}\ {X,Y|X=E[X]}} = P\left[Y > {\gamma_1 \over \gamma_2}\right] & &
\end{eqnarray}

\begin{figure}[htbp]
        \centerline{\psfig{file=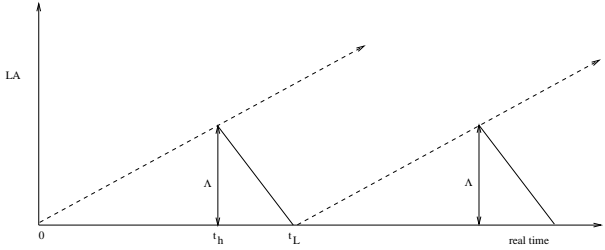,width=3.125in}}
        \caption{Lookahead with a Sliding Lookahead Window.}
        \label{lookahead}
\end{figure}

The three probabilities in Equations \ref{fpcache}
through \ref{fpslow} depend on ($Y$) and real\index{Real} time because
the analysis assumes that the lookahead increases 
indefinitely which shifts the thresholds in such a manner as to 
increase \acl{AVNMP} performance as real\index{Real} time increases. 
However, in this analysis, it is assumed that a sliding lookahead window
exists on each \acl{LP} to control the lookahead rate locally.
The 
\acl{AVNMP} algorithm holds processing of virtual\index{Active Virtual Network Management Protocol!verification} messages once the 
end of the \acf{SLW} is reached. The hold time occurs when $LA = \Lambda$ 
where $\Lambda$ is the length of the \acl{SLW}. Once $\Lambda$ is reached, 
processing of virtual\index{Active Virtual Network Management Protocol!verification} messages is discontinued until real\index{Real}-time reaches 
\acl{LVT}. The lookahead versus real\index{Real} time including the effect of the \acl{SLW} 
is shown in Figure \ref{lookahead}. The dashed arrow represents the 
lookahead which increases at rate $PR$. The solid line returning to
zero is lookahead as the \acl{LP} delays. Because the curve in
Figure \ref{lookahead} from $0$ to $t_L$ repeats indefinitely, only the area
from $0$ to $t_L$ need be considered. For
each $P_i(t)$ $i=1,2,3$, the time average over the lookahead time ($t_L$) 
is shown 
by the integral in Equation \ref{ala}.

\begin{equation}
\label{ala}
P_{X,Y|X=E[X]} = {1 \over t_L} \int_0^{t_L} P_i(t) dt
\end{equation}

\begin{figure}[htbp]
        \centerline{\psfig{file=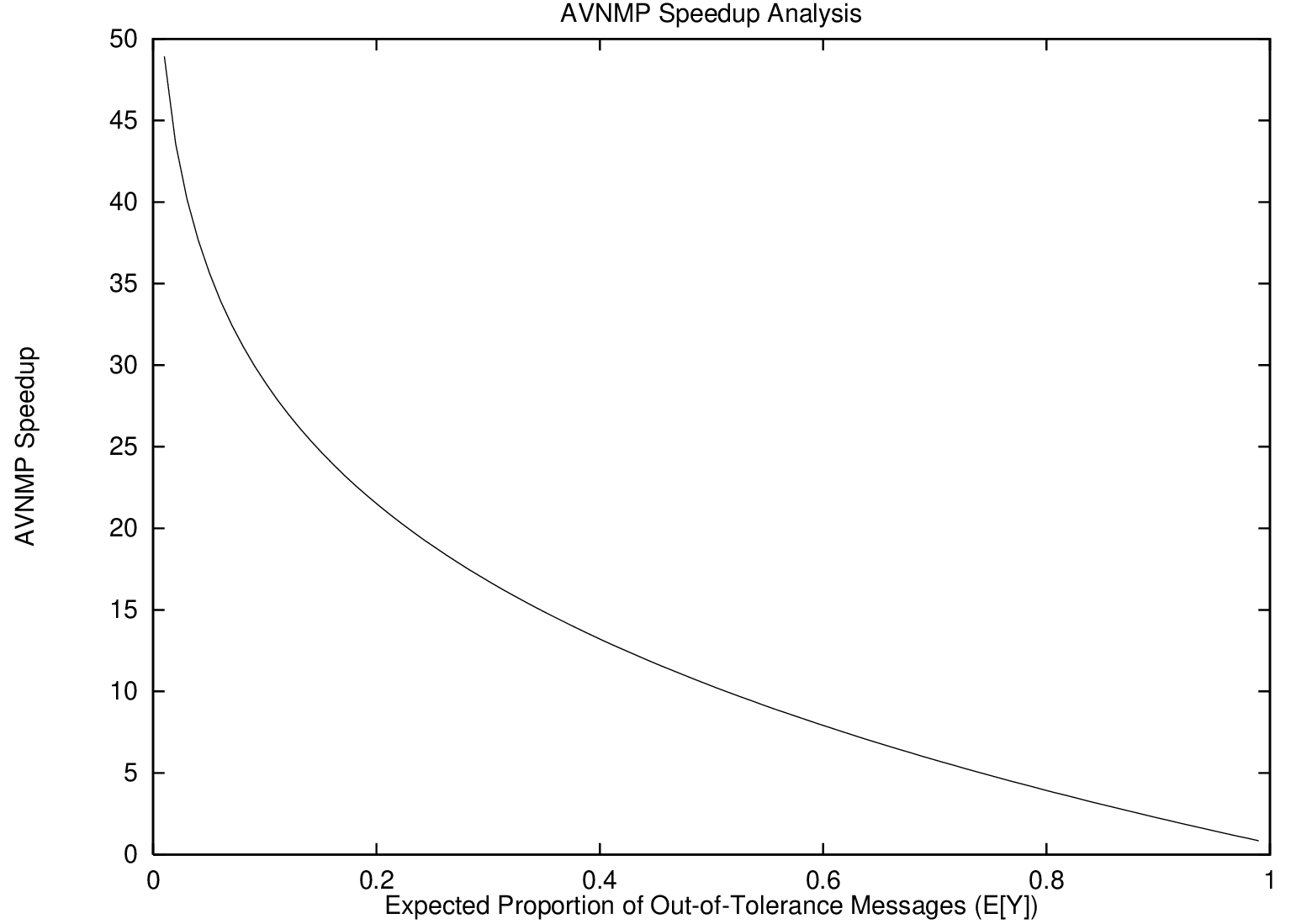,width=3.125in}}
        \caption{AVNMP Speed\index{Speed}up.}
        \label{speedup}
\end{figure}

The probability of each of the events shown in Figures
\ref{cachesu1} through \ref{cachesu3} is multiplied by the speedup\index{Speedup}
for each event in order to derive the average speedup\index{Speedup}. For the 
case
shown in Figure \ref{cachesu1}, the speedup\index{Speedup} ($C_r$) is provided by the time
to read the cache over directly computing the result. For the remaining
cases the speedup\index{Speedup} is $PR_{X,Y|X=E[X]}$ which has been defined as 
${LVT_{X,Y|X=E[X]} \over t}$ as shown in Equation \ref{eta}.
The analytical results for speedup\index{Speedup} are graphed in Figure \ref{speedup}.
A high probability of out-of-tolerance rollback\index{Rollback} in Figure 
\ref{speedup} results in a speedup\index{Speedup} of less than one. Real\index{Real Message} messages
are always processed when they arrive at an \acl{LP}. Thus, no matter
how late \acl{AVNMP} results are, the system will continue to run near
real time.  However, when \acl{AVNMP} results are very late due to a high
proportion of out-of-tolerance messages, the \acl{AVNMP} system is slower than 
real time because out-of-tolerance rollback overhead\index{Active Virtual Network Management Protocol!overhead} processing occurs.
Anti-messages must be sent to correct other \acl{LP}es which
have processed messages which have now been found to be out of tolerance from 
the current \acl{LP}. This causes the speedup\index{Speedup} to be less than one when the 
out-of-tolerance probability is high. Thus, $PR_{X,Y|X=E[X]} < 1$  
for the ``slow'' case shown in Figure \ref{cachesu3}.

\begin{table*}
\begin{eqnarray}
\label{eta}
\lefteqn{\eta \equiv } \\
& & P_{{cache}\ {X|X=E[X]}} C_r + (P_{{late}\ {X|X=E[X]}} + 
P_{{slow}\ {X|X=E[X]}}) PR_{X,Y|X=E[X]} \nonumber
\end{eqnarray}
\end{table*}

\begin{figure}[htbp]
        \centerline{\psfig{file=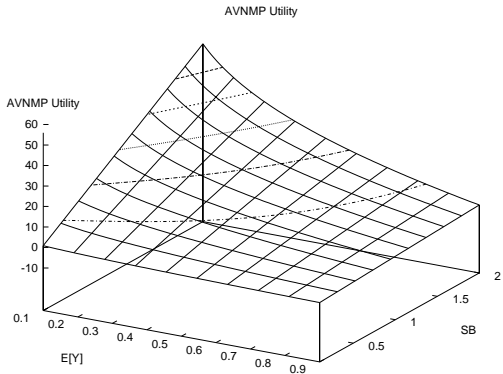,width=3.125in}}
        \caption{Overhead versus Speed\index{Speed}up as a Function of Probability of Rollback\index{Rollback}.}
        \label{overspeed}
\end{figure}

Equation \ref{vncutil1} shows the complete \acl{AVNMP} performance utility.
The surface plot showing the utility of \acl{AVNMP} 
as a function\index{Function} of the proportion of out-of-tolerance 
messages is shown
in Figure \ref{overspeed} where $\Phi_s$, $\Phi_w$, $\Phi_b$ are one.
\aparms. 
The y-axis is the relative marginal utility of speedup\index{Speedup} over reduction
in bandwidth overhead\index{Active Virtual Network Management Protocol!overhead} $SB = \frac{\Phi_s}{\Phi_b}$. Thus if bandwidth 
reduction is much more important than speedup\index{Speedup}, the utility is low and 
the proportion of rollback\index{Rollback} messages would have to be kept below 0.3
in this case. However, if speedup\index{Speedup} is of higher priority
relative to bandwidth, the proportion of out-of-tolerance rollback\index{Rollback} 
message values can be as high as 0.5. If the 
proportion of out-of-tolerance messages becomes too high, the utility 
becomes negative because prediction time begins to fall behind real\index{Real} time.

\begin{figure}[htbp]
        \centerline{\psfig{file=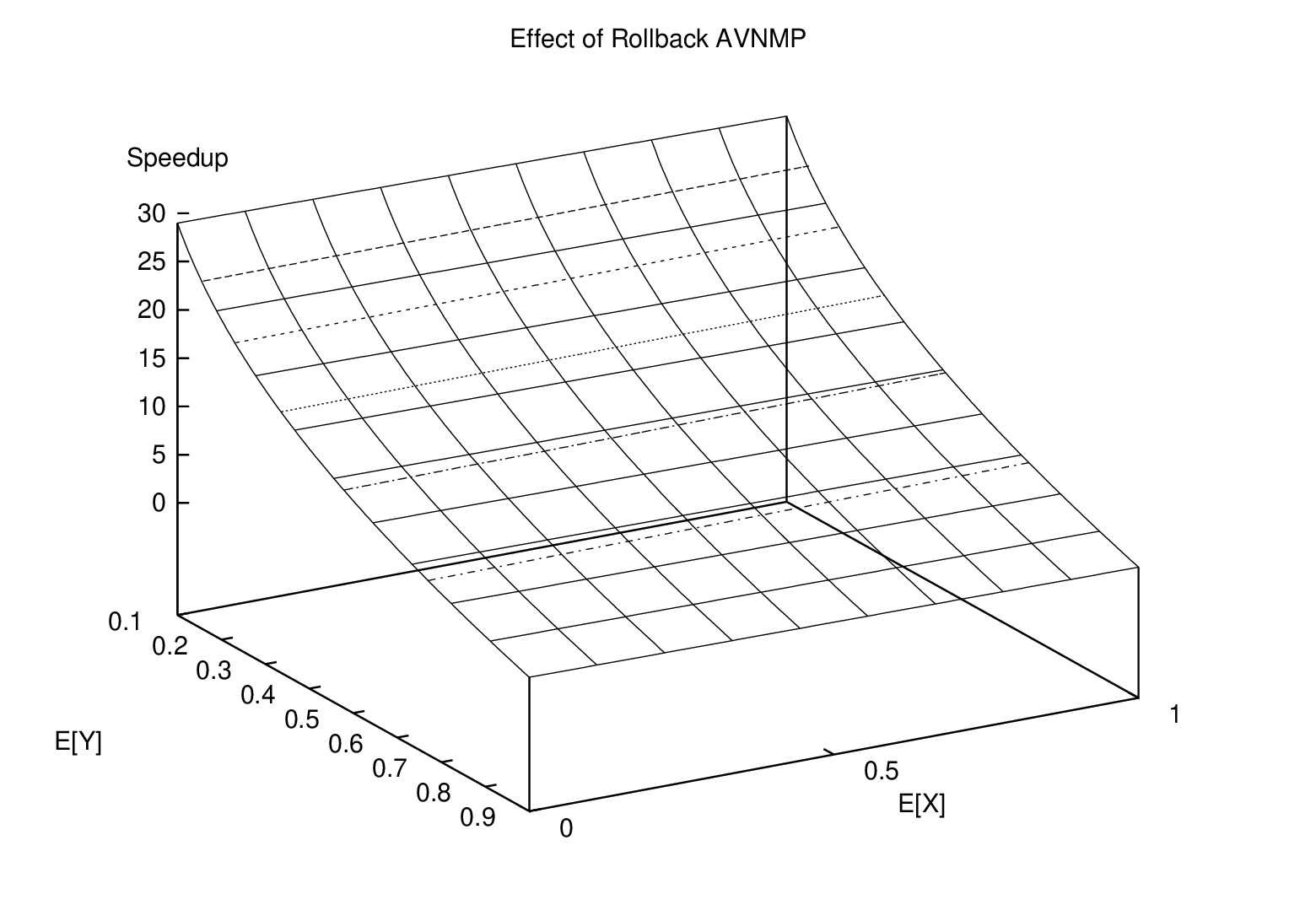,width=3.125in}}
        \caption{Effect of Non-Causality and Tolerance\index{Tolerance} on Speed\index{Speed}up.}
        \label{rollbacks}
\end{figure}

The effect of the proportion of out-of-order and out-of-tolerance 
messages on \acl{AVNMP} speedup\index{Speedup} is shown in Figure \ref{rollbacks}. 
This graph shows that out-of-tolerance rollback\index{Rollback}s have a greater impact 
on speedup\index{Speedup} than out-of-order rollback\index{Rollback}s. The reason for the greater impact 
of the proportion of out-of-tolerance messages is that rollback\index{Rollback}s 
caused by such messages always cause a process to rollback\index{Rollback} to real\index{Real} time. 
An out-of-order rollback\index{Rollback} only requires the process to rollback\index{Rollback} to the 
previous saved state.

\begin{figure}[htbp]
        \centerline{\psfig{file=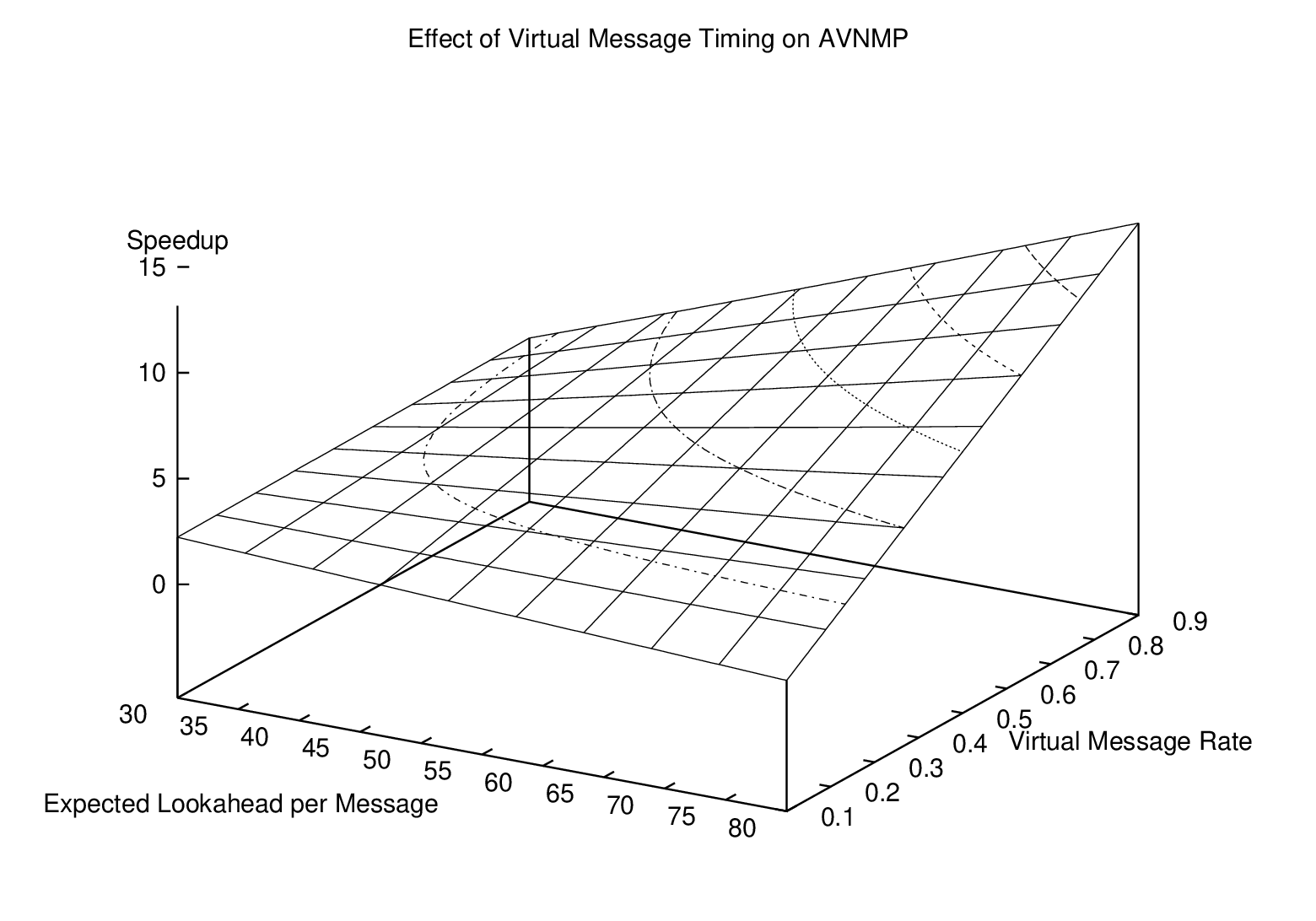,width=3.125in}}
        \caption{Effect of Virtual\index{Active Virtual Network Management Protocol!verification} Message Rate and Lookahead on Speed\index{Speed}up.}
        \label{vmrate}
\end{figure}

Figure \ref{vmrate} shows the effect of the proportion of virtual\index{Active Virtual Network Management Protocol!verification}
messages and expected lookahead per virtual\index{Active Virtual Network Management Protocol!verification} message on speedup\index{Speedup}.
This graph is interesting because it shows how the proportion of virtual\index{Active Virtual Network Management Protocol!verification}
messages injected into the \acl{AVNMP} system and the expected lookahead
time of each message can affect the speedup\index{Speedup}. 
The real\index{Real} and virtual\index{Active Virtual Network Management Protocol!verification} message rates are 0.1 messages per millisecond and the
remaining parameters remain as previously stated.

\begin{figure*}
\begin{eqnarray}
\label{vncutil1}
U_{AVNMP} & = & \left( P_{{cache}\ {X|X=E[X]}} C_r + \right. \\ \nonumber
        &   & \left. (P_{{late}\ {X|X=E[X]}} + P_{{slow}\ {X|X=E[X]}}) PR_{X,Y|X
=E[X]} \right) \Phi_s - \\ \nonumber
        &   & P[|AC_t(\Lambda)| > \Theta] \Phi_w  - \\ \nonumber
        &   & \left({\frac{\lambda_v}{\lambda_{rb}} + \lambda_{v}}
                   \over \lambda_r\right) \Phi_b \nonumber
\end{eqnarray}
\end{figure*}

\Section{Summary}

This paper has introduced a novel algorithm, \acf{AVNMP},
for predictive network management. It explained how 
the \acl{AVNMP} facilitates the management of an active network by 
allowing future predicted state information within an active network 
to be available to network management algorithms. This has been 
accomplished by coupling ideas from optimistic discrete event 
simulation with active networking. The optimistic discrete event 
simulation method used is a form of self-adjusting Time Warp. The
concept of a streptichron and autoanaplasis have been introduced as
mechanisms which take advantage of the enhanced flexibility and intelligence 
of active packets. Finally, it has been demonstrated that the \acl{AVNMP} is 
a feasible concept.

Future work on the \acl{AVNMP} will focus on dynamic message
reprioritization such that both real and virtual messages are processed
with optimal priority as well as automatic dynamic deployment of
the \acl{AVNMP} logical and driving processes within a network. Work 
will also continue on
finding synergies between the self-adjusting Time Warp algorithm and
active networks in order to optimize the \acl{AVNMP}.
%
 \acrodef{A}{Anti-Message}
 \acrodef{ABR}{Available Bit Rate}
 \acrodef{AHDLC}{Adaptive High Level Data Link Protocol}
 \acrodef{AMPS}{Advanced Mobile Phone System}
 \acrodef{ANO}{Attach New to Old}
 \acrodef{ARP}{Address Resolution Protocol}
 \acrodef{ATMARP}{Asynchronous Transfer Mode Address Resolution Protocol}
 \acrodef{ATM}{Asynchronous Transfer Mode}
 \acrodef{AVNMP}{Active Virtual Network Management Protocol}
 \acrodef{BTS}{Base Transceiver Station}
 \acrodef{C/E}{Condition Event Network}
 \acrodef{CDMA}{Code Division Multiple Access}
 \acrodef{CDPD}{Cellular Digital Packet Data}
 \acrodef{CE}{Clustered Environment}
 \acrodef{CMB}{Chandy-Misra-Bryant}
 \acrodef{CMIP}{Common Management Information Protocol}
 \acrodef{CRC}{Cyclic Redundancy Checksum}
 \acrodef{CSS}{Cell Site Switch}
 \acrodef{CS}{Current State}
 \acrodef{CTW}{Clustered Time Warp}
 \acrodef{DES}{Data Encryption Standard}
 \acrodef{DHCP}{Dynamic Host Configuration Protocol}
 \acrodef{DP}{Driving Process}
 \acrodef{EN}{Edge Node}
 \acrodef{ES}{Edge Switch}
 \acrodef{ESN}{Electronic Serial Number}
 \acrodef{FH}{Fixed Host}
 \acrodef{FSM}{Finite State Machine}
 \acrodef{GFC}{Generic Flow Control}
 \acrodef{GPS}{Global Positioning System}
 \acrodef{GSM}{Global System for Mobile Communication}
 \acrodef{GSV}{Global Synchronic Distance}
 \acrodef{GVT}{Global Virtual Time}
 \acrodef{HDLC}{High Level Data Link Protocol}
 \acrodef{HLR}{Home Location Register}
 \acrodef{HO}{Hand-off}
 \acrodef{IETF}{Internet Engineering Task Force}
 \acrodef{IPC}{Inter-Processor Communication}
 \acrodef{LIS}{Logical IP Subnet}
 \acrodef{LN}{Logical Node}
 \acrodef{LP}{Logical Process}
 \acrodef{LVT}{Local Virtual Time}
 \acrodef{MAC}{Media Access Control}
 \acrodef{MH}{Mobile Host}
 \acrodef{MIB}{Management Information Base}
 \acrodef{MIN}{Mobile Identification Number}
 \acrodef{MSC}{Mobile Switching Center}
 \acrodef{MSR}{Mobile Support Router}
 \acrodef{MTW}{Moving Time Windows}
 \acrodef{MT}{Mobile Terminal}
 \acrodef{NCP}{Network Control Protocol}
 \acrodef{NHRP}{Next Hop Resolution Protocol}
 \acrodef{NIPAT}{Network Insecurity Path Assessment Tool}
 \acrodef{NPSI}{Near Perfect State Information}
 \acrodef{NTP}{Network Time Protocol}
 \acrodef{PA}{Perturbation Analysis}
 \acrodef{PBS}{Portable Base Station}
 \acrodef{PCN}{Personal Communications Network}
 \acrodef{PDES}{Parallel Discrete Event Simulation}
 \acrodef{PDU}{Protocol Data Unit}
 \acrodef{PGL}{Peer Group Leader}
 \acrodef{PG}{Peer Group}
 \acrodef{P/T}{Place Transition Net}
 \acrodef{PIPS}{Partially Implemented Performance Specification}
 \acrodef{PNNI}{Private Network-Network Interface}
 \acrodef{PP}{Physical Process}
 \acrodef{Q.2931}{Q.2931}
 \acrodef{QR}{Receive Queue}
 \acrodef{QS}{Send Queue}
 \acrodef{QoS}{Quality of Service}
 \acrodef{RDRN}{Rapidly Deployable Radio Network}
 \acrodef{RN}{Remote Node}
 \acrodef{RT}{Real Time}
 \acrodef{SID}{Station Identification}
 \acrodef{SLP}{Single Processor Logical Process}
 \acrodef{SLW}{Sliding Lookahead Window}
 \acrodef{SNMP}{Simple Network Management Protocol}
 \acrodef{SQ}{State Queue}
 \acrodef{SS7}{Signaling System 7}
 \acrodef{TDMA}{Time Division Multiple Access}
 \acrodef{TDN}{Temporary Directory Number}
 \acrodef{TNC}{Terminal Node Controller}
 \acrodef{TOE}{Time of Expiry}
 \acrodef{TR}{Receive Time}
 \acrodef{TS}{Send Time}
 \acrodef{VC}{Virtual Circuit}
 \acrodef{VCI}{Virtual Circuit Identifier}
 \acrodef{VLR}{Visiting Location Register}
 \acrodef{VNC}{Virtual Network Configuration}
 \acrodef{VP}{Virtual Path}
 \acrodef{VPI}{Virtual Path Identifier}
 \acrodef{VTRP}{Virtual Trees Routing Protocol}

%
%
\medskip
\bibliographystyle{latex8}
\bibliography{/home/bushsf/ref/mob,/home/bushsf/ref/twe,/home/bushsf/ref/psim,/home/bushsf/ref/atmmob,/home/bushsf/ref/standards,/home/bushsf/ref/vci,/home/bushsf/ref/handoff,/home/bushsf/ref/theorem_proving,/home/bushsf/ref/topo,/home/bushsf/ref/signaling,/home/bushsf/ref/an}

\begin{thebibliography}{10}\setlength{\itemsep}{-1ex}\small

\bibitem{BrClSh94}
R.~Braden, D.~Clark, and S.~Shenker.
\newblock Integrated services in the internet architecture: an overview, June
  1994.
\newblock RFC \#1633.

\bibitem{BushThesis}
S.~F. Bush.
\newblock {\em {The Design and Analysis of Virtual Network Configuration for a
  Wireless Mobile ATM Network}}.
\newblock PhD thesis, University of Kansas, Aug. 1997.

\bibitem{BushIEEEBB96}
S.~F. Bush, V.~S. Frost, and J.~B. Evans.
\newblock {Network Management of Predictive Mobile Networks}.
\newblock In {\em IEEE Symposium on Planning and Design of Broadband Networks},
  October 1996.
\newblock Session 8, Montebello, Quebec, Canada.

\bibitem{BushWinet97}
S.~F. Bush, S.~Jagannath, J.~B. Evans, V.~Frost, G.~Minden, and K.~S.
  Shanmugan.
\newblock {A Control and Management Network for Wireless ATM Systems}.
\newblock {\em ACM-Baltzer Wireless Networks (WINET)}, 3:267,283, 1997.
\newblock URL: \htmladdnormallink{http:/\-/\-www.ittc.ukans.edu/\-$\sim$sbush}
  {http:/\-/\-www.ittc.ukans.edu/\-~sbush}.

\bibitem{CMIP}
ISO.
\newblock {Open Systems Interconnection - Management Protocol Specification -
  Part 2: Common Management Information Protocol}.

\bibitem{Jefferson82}
D.~R. Jefferson and H.~A. Sowizral.
\newblock {Fast Concurrent Simulation Using The Time Warp Mechanism, Part {I}:
  Local Control}.
\newblock Technical Report TR-83-204, The Rand Corporation, 1982.

\bibitem{Kulk}
A.~B. Kulkarni, G.~J. Minden, R.~Hill, Y.~Wijata, S.~Sheth, H.~Pindi,
  F.~Wahhab, A.~Gopinath, and A.~Nagarajan.
\newblock {Implementation of a Prototype Active Network}.
\newblock In {\em OPENARCH '98}, 1998.

\bibitem{nips-10:Ma+Ji:1998}
S.~Ma and C.~Ji.
\newblock Wavelet models for video time-series.
\newblock In M.~I. Jordan, M.~J. Kearns, and S.~A. Solla, editors, {\em
  Advances in Neural Information Processing Systems}, volume~10. The {MIT}
  Press, 1998.

\bibitem{SNMP}
M.~T. Rose.
\newblock {\em {The Simple Book, An Introduction to the Management of TCP/IP
  Based Internets}}.
\newblock Prentice Hall, 1991.

\bibitem{RDRN}
R.~Sanchez, J.~Evans, G.~Minden, and V.~S. F. K.~S. Shanmugan.
\newblock {RDRN: A Rapidly Deployable Radio Network - Implementation and
  Experience}.
\newblock In {\em ICUPC '98}, 1998.

\bibitem{TSSWM97}
D.~L. Tennenhouse, J.~M. Smith, W.~D. Sincoskie, D.~J. Wetherall, and G.~J.
  Minden.
\newblock A survey of active network research.
\newblock {\em {IEEE} Communications Magazine}, 35(1):80--86, Jan. 1997.

\end{thebibliography}
\end{document}